\documentclass[aps,prl,twocolumn,groupedaddress,showpacs,floatfix]{revtex4}
\usepackage{graphicx}
\begin{document}
\pagenumbering{arabic}
\pagestyle{plain}
%\preprint{}
\title{$^{77}$Se NMR measurements of the $\pi -d$ exchange field in the organic conductor $\lambda-$(BETS)$_{2}$FeCl$_{4}$}
\author{Guoqing Wu,$^{1}$ W. G. Clark,$^{1}$ S. E. Brown,$^{1}$ J. S. Brooks, $^{2}$ A. Kobayashi,$^{3}$ and H. Kobayashi $^{4}$}
\affiliation{$^{1}$Department of Physics and Astronomy, UCLA, Los Angeles, California 90095, USA}
\affiliation{$^{2}$National High Field Laboratory and Florida State University, Tallahassee, Florida 32306, USA}
\affiliation{$^{3}$Research Center of Spectrochememistry, University of Tokyo, Japan}
\affiliation{$^{4}$Institute of Molecular Science, Okazaki, Japan}
\date{\today }
\begin{abstract}
    $^{77}$Se-NMR spectrum and frequency shift measurements in the paramagnetic metal (PM) and antiferromagnetic insulating (AFI) phases are reported for a small single crystal of the organic conductor $\lambda-$(BETS)$_{2}$FeCl$_{4}$ as a function of temperature ($T$) and field alignment for an applied magnetic field $B_{0}$ = 9 T. The results show that in the low $T$ limit, where the localized Fe$^{3+}$ spins ($S_{d}$ = 5/2) are almost fully polarized, the conduction electrons (Se $\pi$-electrons, spin $s_{\pi}$ = 1/2) in the BETS molecules experience an exchange field ($\bf{B}$$_{\pi d}$) from the Fe$^{3+}$ spins with a value of $-$ 32.7 $\pm$ 1.5 T at 5 K and 9 T aligned opposite to $\bf{B}$$_{0}$. This large negative value of $\bf{B}$$_{\pi d}$ is consistent with that predicted by the resistivity measurements and supports the Jaccarino-Peter internal field-compensation mechanism being responsible for the origin of field-induced superconductivity.
\end{abstract}
\maketitle
\pagenumbering{arabic} \pagestyle{plain}
%\preprint{}
%\section{Introduction}
%Put \label in argument of \section for cross-referencing
     Correlations between conduction electrons and local magnetic moments in condensed matter physics are of considerable interest in situations where the properties of the conduction electrons are significantly tuned by the internal field generated by the local magnetic moments. These interactions can lead to a rich variety of phases, including superconductivity, density waves, and magnetic ordering. Many examples include low-dimensional organic conductors that have been synthesized in recent decades \cite{day, coronado, uji1, kobayashi1}. It is widely accepted that their physical properties are largely determined by the interaction between the donor HOMO (highest occupied molecular orbitals) band molecules and the anions \cite{mori1, ruderman, mori2}. 

     An important example is the quasi-two dimensional (2D) triclinic (space group P$\bar{1}$) salt, $\lambda $-(BETS)$_{2}$FeCl$_{4}$, where BETS is bis(ethylenedithio)tetraselenafulvalene (C$_{10}$S$_{4}$Se$_{4}$H$_{8}$) \cite{uji1, kobayashi1, tokumoto, brossard, akutsu1}. Below an applied magnetic field ($\bf{B}$$_{0}$) of about 11 T, as the temperature ($T$) is lowered it has a transition from a paramagnetic metal (PM) to an antiferromagnetic insulating (AFI) phase. At higher fields and low $T$ there is a PM to field-induced superconducting (FISC) phase \cite{uji1, kobayashi1, tokumoto, brossard}.
\begin{figure}
\includegraphics[scale= 0.21]{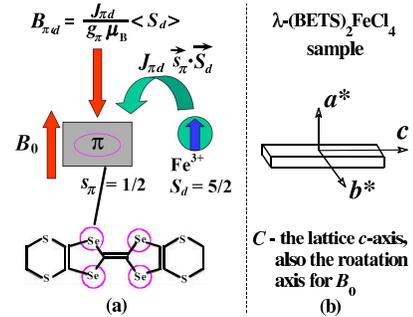}%\\
\caption{(a) Cartoon of the interactions causing the Jaccarino-Peter (JP) compensation mechanism. (b) Sketch of the needle-like shape $\lambda $-(BETS)$_{2}$FeCl$_{4}$ sample for the measurement.\label{fig1}}
\end{figure}

     A mechanism proposed for the FISC phase in $\lambda$-(BETS)$_{2}$FeCl$_{4}$ is the Jaccarino-Peter (J-P) compensation effect \cite{jaccarino} operating in a two-dimensional (2D) system \cite{uji1, balicas1, balicas2}, as illustrated in Fig. 1 (a) \cite{hiraki}. In this model, the total (negative) exchange interaction ($\pi-$d interaction, exchange constant $J_{\pi d}$) between the paramagnetic 3d Fe$^{3+}$ moments ($g\mu_{B}\bf{S}$$_{d}$) ($g$ is the Land\'{e} $g$-factor of the Fe$^{3+}$ and $\mu_{B}$ is the Bohr magneton) and the conduction $\pi$-electrons at the Se sites in the BETS molecule \cite{uji1, akutsu1} generates a large exchange field ($\bf{B}$$_{\pi d}$) at the Se electrons aligned opposite to $\bf{B}$$_{0}$ given by 
\begin{equation}
\bf{B}_{\it{\pi d}} (B_{0}, \it{T}) = \frac{\it{J}_{\pi d}}{\it{g_{\pi}}\mu_{B}}<\bf{S}_{\it{d}} (B_{0}, \it{T})>,
\end{equation}
where $<\bf{S}$$_{d}>$ is the average value of the Fe$^{3+}$ spin polarization, and $g_{\pi}$ is the Land\'{e} $g$-factor for the $\pi$-electrons. When $\bf{B}$$_{0}$ is aligned parallel to the $ac$ plane, the orbital pair breaking effect for the $\pi$-electrons is minimized. At low $T$ ($T$ $<$ 5 K) for 17 T $<$ $\bf{B}$$_{0}$ $<$ 45 T [in the FISC phase] \cite{uji1}, $<\bf{S}$$_{d}>$ is nearly saturated and it is expected that $|$$\bf{B}$$_{\pi d}|$ $\simeq$ 33 T \cite{balicas1}. Thus, the magnitude of the effective field at the Se electrons, $|\bf{B}$$_{0}$ $-$ $\bf{B}$$_{\pi d}|$ = $|\bf{B}$$_{0}|$ $-$ 33 T, is small enough to permit the FISC phase. Also, for $|\bf{B}$$_{0}|$ $<$ $|\bf{B}$$_{\pi d}|$ and $|\bf{B}$$_{0}|$ $>$ $|\bf{B}$$_{\pi d}|$ the spin polarization of the conduction electrons is respectively antiparallel and parallel to $\bf{B}$$_{0}$, a feature that can be probed by the hyperfine frequency shift of the $^{77}$Se NMR signal.

     This description in terms of the J-P mechanism is supported by the fact that its iso-structural nonmagnetic and non-3d-electron analog $\lambda$-(BETS)$_{2}$GaCl$_{4}$ exhibits a behavior \cite{kobayashi2, kobayashi3} that is completely different from that of $\lambda$-(BETS)$_{2}$FeCl$_{4}$. Even though the above model of FISC in $\lambda$-(BETS)$_{2}$FeCl$_{4}$ is widely accepted, it needs further experimental confirmation.

     Nuclear magnetic resonance (NMR) is a versatile local probe that is capable of directly measuring the distribution of internal magnetic field and the electron spin dynamics on the atomic scale. Thus, it can be used as a tool to test the validity of the J-P mechanism for FISC in $\lambda$-(BETS)$_{2}$FeCl$_{4}$. 

     $^{77}$Se-NMR measurments \cite{hiraki1} have been reported for a single crystal of $\lambda$-(BETS)$_{2}$FeCl$_{4}$ with $\bf{B}$$_{0}$ = 14.5 T aligned in the $ac$ plane (PM phase) \cite{uji1}. But the value obtained for $|\bf{B}$$_{\pi d}|$ is 23 T, which is $\sim$ 30$\%$ smaller than the 33 T predicted by the electricial resistivity measurements \cite{uji2, balicas1} and a theoretical estimate \cite{mori2}. Also, these measurements do not include other alignments for $\bf{B}$$_{0}$.

     In this paper, we report $^{77}$Se-NMR spectrum and frequency shift measurements in a single crystal of $\lambda $-(BETS)$_{2}$FeCl$_{4}$, for 2.5 K $\leq$ $T$ $\leq$ 30 K over a range of alignments of $\bf{B}$$_{0}$ = 9 T in the plane $\perp $ to the $c$-axis. At this value of $\bf{B}$$_{0}$, the PM-AFI transition is at 3.5 K. Analysis of these results gives $|\bf{B}$$_{\pi d}|$ = (32.7 $\pm$ 1.5) T aligned opposite to $\bf{B}$$_{0}$ at low $T$ and $B_{0}$ $\geq$ 9 T where the Fe$^{3+}$ magnetization is almost fully saturated (the saturation value of $<S_{d}>$ is slightly less than 2.5 at 9 T due to Fe$^{3+}$ $-$ Fe$^{3+}$ antiferromagnetic interaction \cite{guoqing}), consistent with the predicted value 33 T \cite{uji2, balicas1,mori2}. It supports the J-P compensation as the mechanism for the FISC phase in $\lambda$-(BETS)$_{2}$FeCl$_{4}$. An important input for this work is the Fe$^{3+}$ magnetization that obtained from proton NMR measurements on $\lambda$-(BETS)$_{2}$FeCl$_{4}$ \cite{guoqing}, which are not sensitive to conduction electron contributions.

     The sample used for these measurements was grown using a standard method \cite{kobayashi1} without $^{77}$Se enrichment ($^{77}$Se natural abundance = 7.5$\%$). Its dimensions are $a^{\ast}$ $\times $ $b^{\ast}$ $\times$ $c$ = 0.09 mm $\times$ 0.04 mm $\times$ 0.80 mm [Fig. 1 (b)], corresponding to a mass of $\sim$ 7 $\mu g$ with $\sim$ 2.0 $\times$ 10$^{15}$ $^{77}$Se nuclei. Because of the small number of spins, a small microcoil with a filling factor ($\sim$ 0.4) was used. For most acquisitions, 10$^{4}$-10$^{5}$ averages were used on a time scale of $\sim$ 5 min for 10$^{4}$ averages. The gyromagnetic ratio of $^{77}$Se, $\gamma$ = 8.131 MHz/T, is used for data analysis. The sample and coil were rotated on a goniometer (rotation angle $\phi$) whose rotation axis is along the lattice $c$-axis (needle direction), which is $\perp$ $\bf{B}$$_{0}$. Based on the crystal structure \cite{kobayashi1}, the direction of the Se $p_{z}$ orbital is 76.4$^{\circ}$ from the $c$-axis \cite{guoqing1}. Then, the minimum angle between $p_{z}$ and $\bf{B}$$_{0}$ during the rotation of the goniometer is $\phi_{\text{min}}$ = 13.6$^{\circ}$.

     Figure 2 shows the $^{77}$Se-NMR absorption spectrum ($\chi''$) of $\lambda$-(BETS)$_{2}$FeCl$_{4}$, plotted as a function of the frequency shift $\nu$ $-$ $\nu_{0}$ ($\nu_{0}$ = 72.90 MHz) at a few $T$ from 30 to 2.5 K with $\bf{B}$$_{0}$ = 9 T $\parallel$ $\bf{a}$$'$, where $\nu$ $-$ $\nu_{0}$ has a weak $T$-dependence as discussed in more detail later.
\begin{figure}
\includegraphics[scale= 0.29]{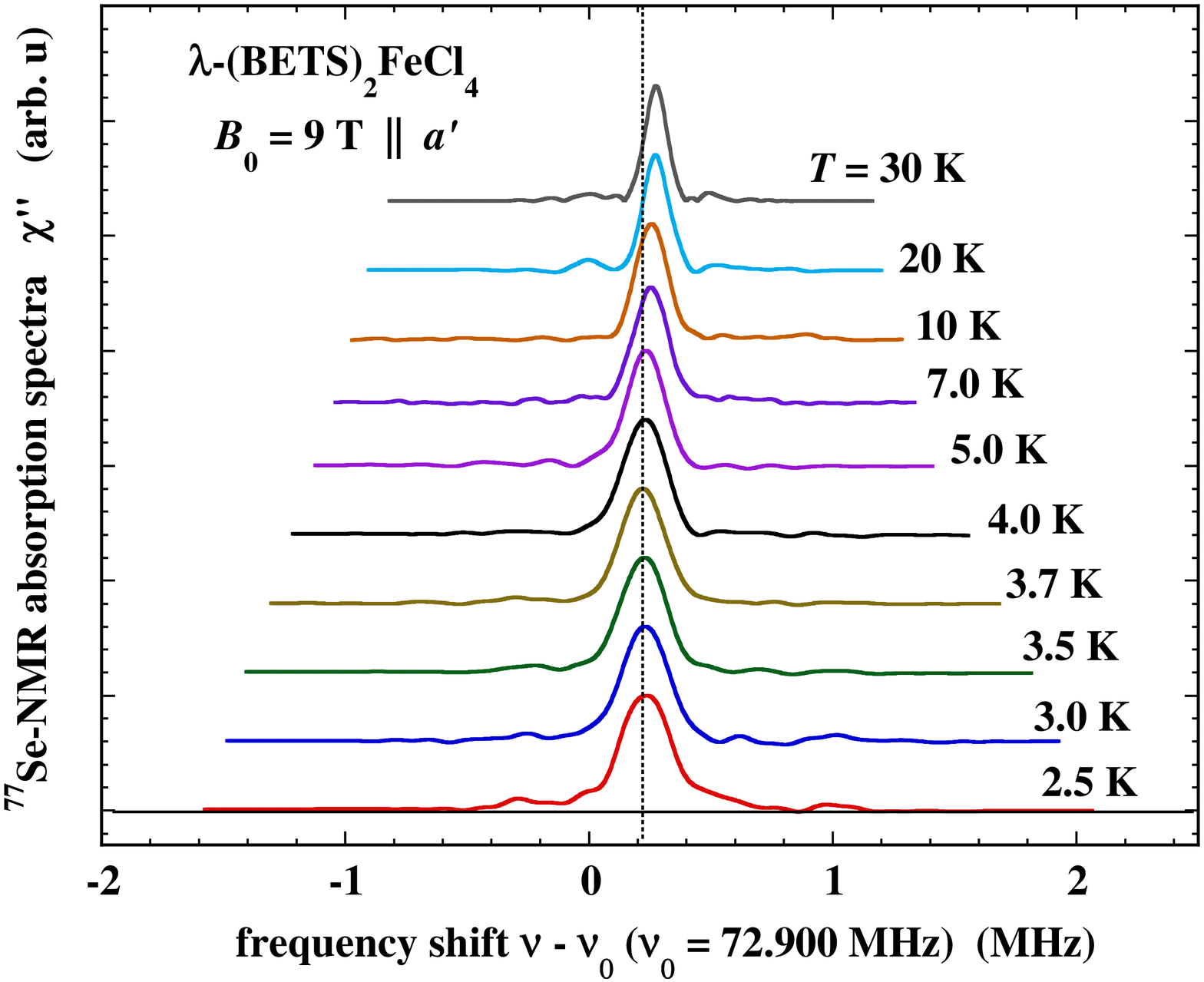}%\\ 
\caption{(Color online) $^{77}$Se-NMR absorption spectrum $\chi''$ as a function of frequency shift at $\nu$ $-$ $\nu_{0}$ ($\nu_{0}$ = 72.90 MHz) at different temperatures with $\bf{B}$$_{0}$ = 9 T $\parallel$ $a'$ in $\lambda$-(BETS)$_{2}$FeCl$_{4}$. \label{fig2}}
\end{figure}

     These spectra measure the distribution of the local magnetic field at the different $^{77}$Se nuclei in the sample. As discussed in more detail later, the local field responsible for this distribution is dominated by the sum of $\bf{B}$$_{0}$, the dipole field from the Fe$^{3+}$ spins ($\bf{B}$$_{\rm{dip}}$), and the hyperfine field from the Se conduction electrons, whose polarization is strongly influenced by $\bf{B}$$_{\pi d}$. The spectra are characterized by (1) the full width at half maximum (FWHM) linewidth ($\Delta f$) which represents the internal magnetic field distribution, and (2) the frequency ($\nu$) of the center of $\chi''$ which measures the average of the hyperfine field from the conduction electrons that coupled to the Fe$^{3+}$ ions. 
\begin{figure}
\includegraphics[scale= 0.29]{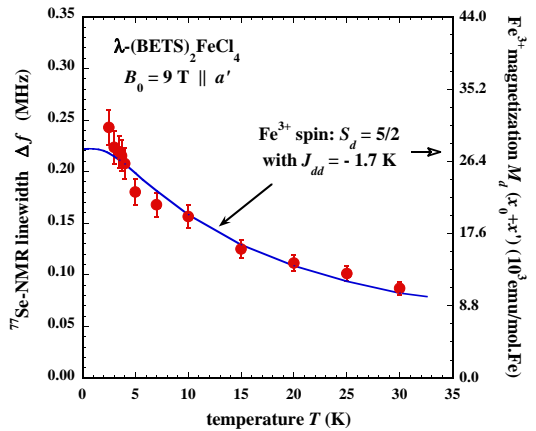}%\\
\caption{(Color online) $T-$dependence of the $^{77}$Se-NMR linewidth (FWHM) $\Delta f$ (solid red circles) and the modified Brillouin function fit of the Fe$^{3+}$ magnetization \cite{guoqing} $M_{d}(x_{0}+x')$ (solid blue line) of $\lambda$-(BETS)$_{2}$FeCl$_{4}$ at $\bf{B}$$_{0}$ = 9 T $\parallel$ $a'$. The error bars are our estimated uncertainty. \label{fig3}}
\end{figure} 

      Figure 3 shows $\Delta f$ as a function of $T$ for $\bf{B}$$_{0}$ = 9 T $\parallel $ $a'$. Also shown is a fit to the Fe$^{3+}$ magnetization $M_{d}(x_{0}+x')$ [unit: 10$^{3}$ emu/mol.Fe] provided by the modified Brillouin function \cite{guoqing} $B_{JM}(x)$ = $B_{J}(x_{0}+x')$, where $B_{J}(x)$ is the standard Brillouin function \cite{ashcroft}, $J$ = $S_{d}$ = 5/2, $x_{0}$ = $\frac{Jg\mu_{B}B_{0}}{k_{B}T}$, $x'$ = $-\frac{Jg\mu _{B}B'}{k_{B}T}$, the $d$-$d$ Fe$^{3+}$ exchange field $B'$ $\simeq$ $\frac{|J_{dd}|Jk_{\rm{B}}}{g\mu _{\rm{B}}}B_{J}(x_{0})$, and $J_{dd}$ = $-$1.7 K is the $d$-$d$ exchange parameter. The fit parameters for $B_{JM}(x)$ are obtained from proton NMR measurements \cite{guoqing}. In Fig. 3, $\Delta f(T)$ increases from $\sim$ 90 kHz to $\sim$ 200 kHz as $T$ is lowered from 30 to 5 K. Also, $M_{d}(x_{0}+x')$ provides a good fit. Since the susceptibility of the BETS molecules is small and nearly independent of $T$, $M_{d}$ is the main source of $\Delta f(T)$ in the PM phase. The size of $\Delta f(T)$ is attributed to a distribution of $\bf{B}$$_{\rm{dip}}$ and the hyperfine field of the Fe$^{3+}$ across the different Se sites. Since measurements of the $^{77}$Se-NMR spin-echo decay indicate a homogeneous linewidth of $\sim$ 10 kHz \cite{guoqing}, it follows that $\chi''$ is strongly inhomogeneously broadened.

       A similar $T-$dependence is also observed in $\nu$, which is plotted as a function of $T$ in Fig. 4 (a) and $M_{d}(x_{0} + x')$ in Fig. 4 (b). In the PM state above $\sim$ 7 K, a good fit to $\nu$ (uncertainty $\pm$ 3 kHz) is obtained using 
\begin{equation}
\nu \simeq [73.221-3.0158\times 10^{-3}M_{d}(x_{0}+x^{\prime })] ~~\rm{MHz}.
\end{equation}

      This result is a strong indication that the $T-$dependence of $\nu$ is dominated by the hyperfine field from the Fe$^{3+}$ magnetization. The negative sign of the contribution from $M_{d}$ is very important. It indicates that the hyperfine field from the Fe$^{3+}$ magnetization is negative, i.e., opposite to $\bf{B}$$_{0}$, as needed for the J-P comensation mechanism.  
\begin{figure}
\includegraphics[scale= 0.32]{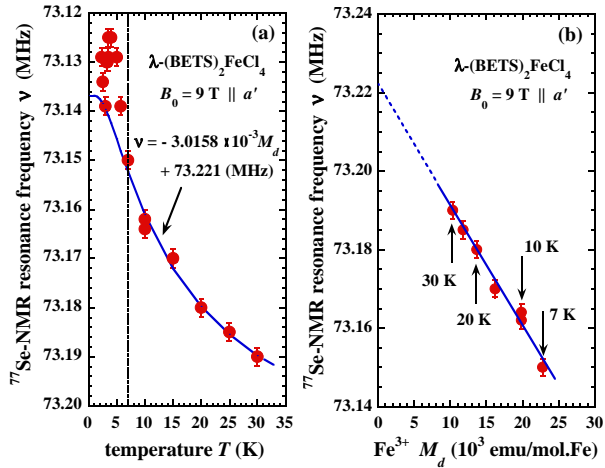}%\\
\caption{(Color online) (a) $T-$dependence of the $^{77}$Se-NMR resonance frequency $\nu$ (solid red) for $\bf{B}$$_{0}$ = 9 T $\parallel$ $a'$ in $\lambda$-(BETS)$_{2}$FeCl$_{4}$. The solid line is a fit to the modified Brillouin function Fe$^{3+}$ magnetization \cite{guoqing} $M_{d}(x_{0}+x')$. (b) $\nu $ vs $M_{d}(x_{0}+x')$ above 7 K for $\bf{B}$$_{0}$ = 9 T $\parallel$ $a'$ in $\lambda$-(BETS)$_{2}$FeCl$_{4}$. The error bars are our estimated uncertainty.}
\label{fig4}
\end{figure}

      A more informative result is shown in Fig. 5, where $\nu$ $-$ $\nu_{0}$ ($\nu_{0}$ = 72.90 MHz) is plotted as a function of both $\phi$ and $T$. As shown in the lower right of Fig. 5, the $z$-axis is chosen to be parallel to the $c$-axis, $p_{z}$ is the $z$-component of the BETS $\pi $-electron orbital moment, the $x$-axis is in the the $c$-$p_{z}$ plane, and $\bf{B}$$_{0}$ is in the $xy$-plane ($\perp$ $c$) rotated by $\phi$ from the $x$-axis. The angle $\phi$ = 0$^{\circ }$ corresponds to $\bf{B}$$_{0}$ $\parallel$ the $c$-$p_{z}$ plane and the minimum angle between $\bf{B}$$_{0}$ and $p_{z}$ is $\phi_{\rm{min}}$ = 13.6$^{\circ }$. The solid lines are a fit to the following theoretical model based upon the hyperfine coupling to the BETS Se $\pi$-electrons whose polarization is affected by the exchange field from the Fe$^{3+}$ magnetization.

      According to NMR theory \cite{slichter}, the contributions to the Hamiltonian ($H_{I}$) of the $^{77}$Se nuclear spins can be expressed as 
\begin{equation}
H_{I} = H_{IZ} + H_{II} + H_{I\pi }^{\rm{hf}} + H_{Id}^{\rm{hf}} + H_{d}^{\rm{dip}},
\end{equation}
where $H_{IZ}$ is the Zeeman Hamiltonian of the $^{77}$Se nuclei in $\bf{B}$$_{0}$, $H_{II}$ is the $^{77}Se-^{77}$Se dipole-dipole interaction Hamiltonian, $H_{I\pi }^{\rm{hf}}$ is the direct hyperfine coupling of the $^{77}$Se nucleus to the BETS $\pi$-electrons generated by $\bf{B}$$_{0}$, $H_{Id}^{\rm{hf}}$ is the indirect hyperfine coupling via the $\pi$-electrons to the 3d Fe$^{3+}$ spins (field $\bf{B}$$_{\pi d}$), and $H_{d}^{\rm{dip}}$ is the dipolar coupling to the Fe$^{3+}$ which gives $\bf{B}$$_{\rm{dip}}$. Here, $\bf{B}$$_{\rm{dip}}$ is calculated \cite{guoqing1} using the sum of the near dipole, the bulk demagnetization and the Lorentz contributions \cite{slichter,carter}. All of these terms contribute to the static local magnetic field at the $^{77}$Se sites and all but the first cause the $^{77}$Se-NMR frequency shifts. 
\begin{figure}
\includegraphics[scale= 0.32]{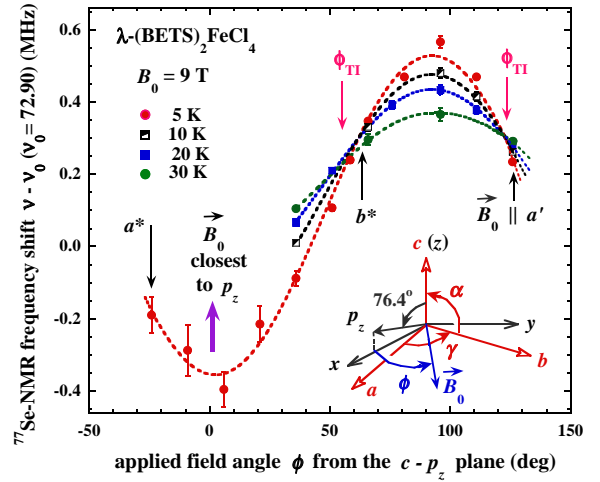}%\\
\caption{(Color online) $^{77}$Se-NMR resonance frequency $\nu$ as a function of angle $\phi$ at several $T$ for the rotation of $B_{0}$ = 9 T about the $c$-axis in $\lambda$-(BETS)$_{2}$FeCl$_{4}$. The error bars are our estimated uncertainty.}
\label{fig5}
\end{figure}

     Based upon the lattice structure and the shape and size of the sample, we calculated the angular dependence of $\bf{B}$$_{\rm{dip}}$ and estimated \cite{guoqing1} that the contributions of $H_{II}$ is negligible. Thus,
\begin{equation}
H_{I}\simeq H_{IZ}+H_{I\pi }^{\rm{hf}}+H_{Id}^{\rm{hf}} + H_{d}^{\rm{dip}},
\end{equation}
and the corresponding value of $\nu$ is \cite{hiraki1}
\begin{eqnarray}
\frac{\nu(\phi^{\prime},B_{0},T)}{\gamma} & = & [B_{0} + B_{\rm{dip}}(B_{0},T)][1 + K_{C} + K_{s}(\phi^{\prime})] \nonumber \\
& & +  K_{s}(\phi^{\prime})B_{\pi d}(B_{0},T) \\
& \simeq & B_{0} (1 + K_{C}) + B_{\rm{dip}}(B_{0},T)  \nonumber\\
& & +  K_{s}(\phi^{\prime})[B_{0} + B_{\pi d}(B_{0},T)] 
\end{eqnarray}
where $\phi^{\prime}$ is the angle between $\bf{B}$$_{0}$ and the $p_{z}$ direction, and $K_{c}$ and $K_{s}$ are respectively the (orbital) chemical shift and the (spin) Knight shift of the BETS Se $\pi$-electrons. The approximation in the second line corresponds to dropping the terms $B_{\rm{dip}}(B_{0},T)[K_{C} + K_{S}(\phi^{\prime})]$, which are the product of small quantities. It can be shown \cite{hiraki1, metzger, takagi, guoqing1} that for rotation of the sample about the $c$-axis with $\bf{B}$$_{0}$ in the plane $\perp$ $c$, $K_{s}(\phi^{\prime})$ is given by
\begin{eqnarray}
K_{s}(\phi^{\prime}) & = & K_{\rm{iso}} + K_{\rm{an}}(\phi^{\prime}) \\
K_{s}(\phi)~ & = & K_{\rm{iso}} + K_{\rm{ax}}[3\cos^{2}\phi\cos^{2}\phi_{\rm{min}}-1],
\end{eqnarray}
where $K_{\rm{iso}}$ and $K_{\rm{an}}(\phi)$ are the isotropic (independent of $\phi$) and axial (anisotropic) parts of the Knight shift, respectively. $K_{\rm{iso}(\rm{ax})}$ is a constant determined by the isotropic (axial) hyperfine field $[A_{\rm{iso}(\rm{ax})}]$ produced by the 4$p_{\pi}$ spin polarization of the BETS Se $\pi$-electrons \cite{hiraki1}.

      The quantities that determine $K_{\rm{ax}}$ are
\begin{eqnarray}
K_{\rm{ax}} & = & \frac{A_{\rm{ax}}}{N_{A}g_{\pi}\mu_{B}}~\chi_{\rm{BETS}} \\
\rm{and}~~~~~~~~~~~~~~~ A_{\rm{ax}} & = & \frac{2}{5}<r^{-3}>_{4p}\mu_{B}\sigma_{\rm{Se}}, ~~~~~~~~~~~~~~~~
\end{eqnarray}
where $\chi_{\rm{BETS}}$ = 4.5 $\times$ 10$^{-4}$ emu/mol is the BETS $\pi$-electron susceptibility \cite{tanaka}, $g_{\pi}$ = 2, $N_{A}$ is the Avogadro's number, and $A_{\rm{ax}}$ = + 38.6 kOe/$\mu_{B}$ reported by S. Takagi $et~ al.$ \cite{takagi}. This value of $A_{\rm{ax}}$ is based on theoretical calculations $<r^{-3}>_{4p}$ = 9.28 $a_{0}^{-3}$ ($a_{0}$ = 0.529 $\AA$, the Bohr radius) \cite{fraga}, and $\sigma_{\rm{Se}}$ = 0.166 \cite{takagi1} obtained for $\lambda$-(BETS)$_{2}$GaCl$_{4}$, which has essentially the same BETS-molecules as $\lambda$-(BETS)$_{2}$FeCl$_{4}$. By using these values, one obtains $K_{\rm{ax}}$ = 15.3 $\times$ 10$^{-4}$. Its uncertainty is not known to us and is not included in our analysis.

      The angular dependence of the shift in $K_{s}(\phi)$ [Eqs. (7)-(8)] has been used for the fit of the 5 K data shown in Fig. 5 using the relation $\nu$ $-$ $\nu_{0}$ = $a_{1}[3\cos^{2}\phi\cos^{2} 13.6^{\circ}$ $-$ 1] + $b_{1}$, with the fit values $a_{1}$ = $-$ 313 kHz and $b_{1}$ = + 221 kHz (uncertainty $\pm$ 8 kHz).

      From Eqs. (6)$-$(8), the formula for $B_{\pi, d}$ at a fixed $B_{0}$ and $T_{0}$ from the difference in $\nu$ ($\Delta\nu$) at the two angles $\phi_{1}$ and $\phi_{2}$ is
\begin{equation}
B_{\pi d}(B_{0},T_{0}) = \frac{\Delta\nu (T_{0},\phi_{1},\phi_{2})}{\gamma\Delta K_{\rm{an}} (\phi_{1},\phi_{2})} - B_{0} - \frac{B_{\rm{dip}}(\phi_{1}) - B_{\rm{dip}}(\phi_{2})}{\Delta K_{\rm{an}}(\phi_{1},\phi_{2})},
\end{equation}
where
\begin{equation} 
\Delta K_{\rm{an}}(\phi_{1},\phi_{2}) = 3K_{\rm{ax}}\cos^{2}\phi_{\rm{min}}(\cos^{2}\phi_{1} - \cos^{2}\phi_{2}).
\end{equation}

      The conditions $T$ = 5 K, $\phi_{1}$ = 90$^{\circ}$, and $\phi_{2}$ = 0$^{\circ}$ give $\Delta\nu$(5K, 90$^{\circ}$, 0$^{\circ}$) = 880 $\pm$ 26 kHz (from Fig. 5) and $\Delta K_{\rm{an}}$ = 4.42 $\times$ 10$^{-3}$ (estimated error $\pm$ 3$\%$). Our calculated value of $B_{\rm{dip}}(0^{\circ})$ $-$ $B_{\rm{dip}}$(90$^{\circ}$) at 5 K is 3.63$\times$10$^{-3}$ T and the field used was $B_{0}$ = 9.0006 T. These values then give $B_{\it{\pi d}}$ = $-$ 32.7 $\pm$ 1.5 T at 5 K (aligned opposite to $B_{0}$ = 9.0006 T), which is very close to the expected value of $-$ 33 T obtained from the electrical resistivity measurement \cite{uji2, balicas1} and the theoretical estimate \cite{mori2}. Also, a small increase in $<\bf{S}$$_{d}>$ is expected as $B_{0}$ is increased from 9 T to 33 T at 5 K. From the modified Brillouin function analysis used for the proton NMR linewidth \cite{guoqing}, we expect this increase in $<\bf{S}$$_{d}>$ to be a factor 1.05 $\pm$ 0.05, which corresponds to an adjustment to $B_{\pi d}(33$ T, 5 K$)$ = $-$ 34.3 $\pm $ 2.4 T.

      Similar results for the exchange field $B_{\pi d}$ can also be obtained from the $T$-dependence of $\nu$ at a fixed $\phi$ with the data in Figs. 4 and 5 using the same kind of fit as Eq. (2). But the uncertainty in the value obtained for $B_{\pi d}$ with this type of analysis is large enough that we do not present it here. An important test we plan to do in the future is to extend the measurements to $B_{0}$ $>$ $|B_{\pi d}|$ to find the value of $B_{0}$ = $|B_{\pi d}|$, where the angular dependence in Fig. 5 disappears [Eqs. (6)$-$(8)] and above which the sign of $K_{s}(\phi^{\prime})$ changes from negative to positive.

      Figure 5 shows two values of $\phi$ ($\phi_{\rm{TI1}}$ and $\phi_{\rm{TI2}}$) where $\nu$ becomes independent of $T$. The measured difference $\phi_{\rm{TI1}}$ $-$ $\phi_{\rm{TI2}}$ = 65.0$^{\circ}$ $\pm$ 2.0$^{\circ}$, which should be symmetric around 90$^{\circ}$, or $\phi_{\rm{TI1}}$ = 57.5$^{\circ}$ $\pm$ 1.0$^{\circ}$ and $\phi_{\rm{TI2}}$ = 122.5$^{\circ}$ $\pm$ 1.0$^{\circ}$. By using Eqs. (6) and (8) and neglecting the very small contribution from $B_{\rm{dip}}$ ($B_{\rm{dip}}$ $<<$ $|B_{\pi d}|$), it can be shown that the condition for $\phi_{\rm{TI\it{i}}}$ ($i$ = 1 , 2) is $K_{\rm{iso}}$ + $K_{\rm{ax}}[3\cos^{2}\phi_{\rm{TIi}}\cos^{2}\phi_{\rm{min}}$ $-$ 1] = 0. From this relation, the measured values of $\phi_{\rm{TI1}}$ and $\phi_{\rm{TI2}}$, and $K_{\rm{ax}}$ = 15.3 $\times$ 10$^{-4}$, one obtains $K_{\rm{iso}}$ = (2.8 $\pm$ 0.7) $\times$ 10$^{-4}$, or $K_{\rm{iso}}/K_{\rm{ax}}$ = 1/(5.5 $\pm$ 1.4).

      In summary, our results of $^{77}$Se-NMR spectrum and frequency shift measurements in $\lambda-$(BETS)$_{2}$FeCl$_{4}$ indicate that the Fe$^{3+}$ spins have a strong antiferromagnetic coupling to the BETS $\pi$-electrons, and we determined the $\pi - d$ exchange field to be $B_{\pi d}$ = $J_{\pi d}<S_{d}>/g\mu_{B}$ = $-$ 32.7 $\pm$ 1.5 T at $B_{0}$ = 9 T and 5 K, with an expected value of $-$ 34.3 $\pm$ 2.4 T at $B_{0}$ $\simeq $ 33 T and 5 K. This large negative value of $B_{\pi d}$ (or $J_{\pi d}$) is consistent with that predicted by the resistivity measurements, and supports the Jaccarino-Peter internal field-compensation mechanism being responsible for the origin of the FISC phase in $\lambda-$(BETS)$_{2}$FeCl$_{4}$.

      This work at UCLA is supported by NSF Grants DMR-0334869 (W.G.C.) and 0520552 (S.E.B.), and work at NHMFL is supported by NSF under Cooperative Agreement No. DMR-0084173 and the State of Florida.

\end{document}